\documentclass[fleqn,usenatbib]{mnras}

\usepackage{newtxtext,newtxmath}
\usepackage[T1]{fontenc}

\DeclareRobustCommand{\VAN}[3]{#2}
\let\VANthebibliography\thebibliography
\def\thebibliography{\DeclareRobustCommand{\VAN}[3]{##3}\VANthebibliography}

\usepackage{graphicx}
\usepackage{amsmath}
\usepackage{booktabs}
\makeatletter
\title[Stellar J-Harvesting: a new technosignature]{Stellar J-Harvesting: a novel angular momentum technosignature and first search in the Kepler field}

\author[S. Torlakcik]{
Sahin Torlakcik$^{1}$\thanks{E-mail: torlakciksahin@gmail.com}
\\
$^{1}$Ankara Atat\"urk High School, Ankara, T\"urkiye
}

\date{Preprint, July 2026}

\pubyear{2026}
\begin{document}
\label{firstpage}
\pagerange{\pageref{firstpage}--\pageref{lastpage}}
\maketitle
\begin{abstract}
We introduce \emph{Stellar J-Harvesting} (SJH), a technosignature in
which an engineered system extracts rotational angular momentum from a
star. Unlike a classical Dyson sphere, such a system need not produce a
detectable mid-infrared excess; its clearest signature would be a star
rotating more slowly than comparable peers. We derive the energy--period
relation for this effect, outline several possible coupling channels, and
apply a population-relative slow-rotator search to Kepler stars using
colour--gravity bins. After eight false-positive filters, the clean sample
contains 6{,}725 FGK main-sequence stars. We find two $>4\sigma$ slow
rotators, but Gaia DR3 indicators and WISE imaging point to more ordinary
explanations, including unresolved binaries and low metallicity. We
therefore make no detection claim and place a conservative pilot upper
limit on the occurrence of strong SJH-like signals,
$f_{\rm SJH}<4.5\times10^{-4}$. The main result is a search framework: angular-momentum technosignatures
are testable with existing stellar-rotation catalogues, and the strongest
outliers define concrete targets for spectroscopic, imaging, and radio
follow-up.
\end{abstract}

\section{Introduction}
\label{sec:intro}

Among technosignature concepts, the Dyson sphere \citep{Dyson1960} has
been the most thoroughly explored in the infrared: a civilization that
captures a significant fraction of its star's luminosity ($L_\star$)
inevitably re-radiates the waste heat as mid-infrared excess, producing
a signature that surveys from IRAS \citep{Carrigan2009} to WISE
\citep{Wright2014} and Gaia \citep{Suazo2024} have pursued with
increasing sensitivity. Yet no confirmed Dyson sphere candidate has
emerged, and the resulting silence echoes Hart's \citep{Hart1975}
formulation of the Fermi paradox.

But luminosity is not the only energy reservoir a star offers. Its
rotational angular momentum ($J_\star$) represents a substantial, yet
entirely unexplored, technosignature target. The Sun's rotational
kinetic energy,
\begin{equation}
E_{\rm rot,\odot} = \tfrac{1}{2} k M_\odot R_\odot^2 \Omega_\odot^2
\approx 2.3 \times 10^{35}\,\mathrm{J},
\end{equation}
where $k \approx 0.059$ is the dimensionless moment of inertia
coefficient, is modest by astrophysical standards. Yet it comes with a
remarkable engineering advantage: accessing it requires far less
material than blocking a star's light, and the waste heat it produces is
so faint that it would be invisible to every existing infrared survey.

We define Stellar J-Harvesting (SJH) as the deliberate extraction
of stellar angular momentum by an engineered structure. The observable
is one-sided: the star rotates slower than its stellar population,
\[
\Delta \log P \equiv \log_{10}\left(\frac{P_{\rm obs}}{P_{\rm pop}}\right) > 0.
\]
This is not a standalone detection criterion, because slow rotation has
natural false-positive channels, but it defines a measurable target class
for follow-up.
Fast rotation can often be explained by youth. Slow rotation, once colour
and evolutionary state are controlled for, is a useful outlier signal to
test.
This concept is distinct from starlifting, which primarily targets
stellar \emph{mass} ($M_\star$) extraction with angular momentum loss
as a byproduct. SJH, Dyson spheres, and starlifting thus define three
orthogonal axes of a technosignature space: $L$-harvesting,
$J$-harvesting, and $M$-harvesting.

Section~\ref{sec:concept} develops the SJH concept and its physical
basis. Section~\ref{sec:method} describes the detection method.
Section~\ref{sec:results} presents the search results.
Section~\ref{sec:discussion} discusses natural alternatives and
limitations.
\vspace{-18pt}
\section{The SJH concept}
\label{sec:concept}

\subsection{How much energy, and what does it look like?}

Because $E_{\rm rot} \propto \Omega^2 \propto P_{\rm rot}^{-2}$,
extracting a fraction $f$ of the rotational kinetic energy yields a
clean, predictable period change:
\begin{equation}
\frac{P_{\rm new}}{P_{\rm old}} = (1-f)^{-1/2}, \quad
\Delta\log P = -\tfrac{1}{2}\log_{10}(1-f).
\end{equation}
Table~\ref{tab:energy} translates this relation into observable
quantities. The median cell-to-cell scatter in our 2D bins is
$\sigma_{\log P} \approx 0.20$\,dex across 24 valid cells (each with
$N \ge 10$ stars) in the 6{,}725-star clean sample. This sets a
$1\sigma$ detection floor at $\Delta\log P \gtrsim 0.20$ ($\gtrsim 60\%$
energy extraction); a $3\sigma$ detection requires $\gtrsim 94\%$
extraction. SJH is therefore detectable only for aggressive, sustained
operations over Gyr timescales at $\dot{E} \gtrsim 10^{18}$\,W.
Outlier significances in Section~\ref{sec:results} use each star's own
cell scatter, $\sigma_{\log P,\rm cell}$, which ranges from 0.16 to
0.35\,dex; the global median serves only as a representative floor.

\begin{table}
\centering
\caption{Energy extraction fraction $f$ versus period anomaly.}
\label{tab:energy}
\begin{tabular}{rrr}
\toprule
$f$ (\%) & $P_{\rm new}/P_{\rm old}$ & $\Delta\log P$ \\
\midrule
50 & 1.41 & 0.15 \\
75 & 2.00 & 0.30 \\
90 & 3.16 & 0.50 \\
95 & 4.47 & 0.65 \\
98 & 7.07 & 0.85 \\
99 & 10.00 & 1.00 \\
\bottomrule
\end{tabular}
\end{table}
\subsection{Why harvest J instead of L?}

Why consider angular momentum when stellar luminosity is $\sim 10^8$ times
larger as an energy reservoir? The motivation is not total energy yield, but
detectability and architecture.

First, \emph{material scale}. The Sun's rotational angular momentum is
\[
J_\odot = k M_\odot R_\odot^2 \Omega_\odot
\approx 1.6 \times 10^{41}\,{\rm kg\,m^2\,s^{-1}} .
\]
Spinning it down over 1\,Gyr requires a mean torque
\[
\tau \sim J_\odot/T \approx 5 \times 10^{24}\,{\rm N\,m}.
\]
As an illustrative estimate, a superconducting loop of radius 1\,AU carrying
current $I$ in the solar-wind magnetic field
($B \sim 10^{-9}$\,T at 1\,AU) has magnetic moment
$m = I\pi R^2$ and torque $\tau \sim mB$, giving
$I \sim 7 \times 10^{10}$\,A. For critical current densities of order
$10^{10}$\,A\,m$^{-2}$, this corresponds to a conductor cross-section of
order a few m$^2$, before allowing for structural support, thermal control,
stability, and deployment losses. The resulting mass scale is therefore at
least plausibly below that of a full luminosity-intercepting Dyson swarm,
which must cover a large fraction of $4\pi R^2$ at AU scales. This comparison
should be read as an order-of-magnitude plausibility argument, not as an
engineering design.

Second, \emph{observational stealth}. Extracting 50\% of
$E_{\rm rot,\odot}$ over 1\,Gyr gives a mean power
$\dot{E} \sim 10^{18}$\,W, about $10^8$ times fainter than the solar
luminosity. At a typical Kepler distance of 1\,kpc, the corresponding flux is
$\sim 10^{-22}$\,W\,m$^{-2}$, well below the WISE W4 5$\sigma$ sensitivity
to mid-infrared excess. SJH would therefore not necessarily resemble a
classical Dyson-sphere technosignature: its primary observable would be a
rotational anomaly, not waste heat.

\subsection{Engineering methods}

The following mechanisms are not proposed as detailed engineering designs.
They are schematic routes showing how angular momentum extraction could, in
principle, couple to a star. In all cases, thermodynamics requires waste heat,
but the relevant power scale can remain far below the luminosity intercepted
by a Dyson swarm.

\begin{figure*}
\centering
\includegraphics[width=0.95\textwidth]{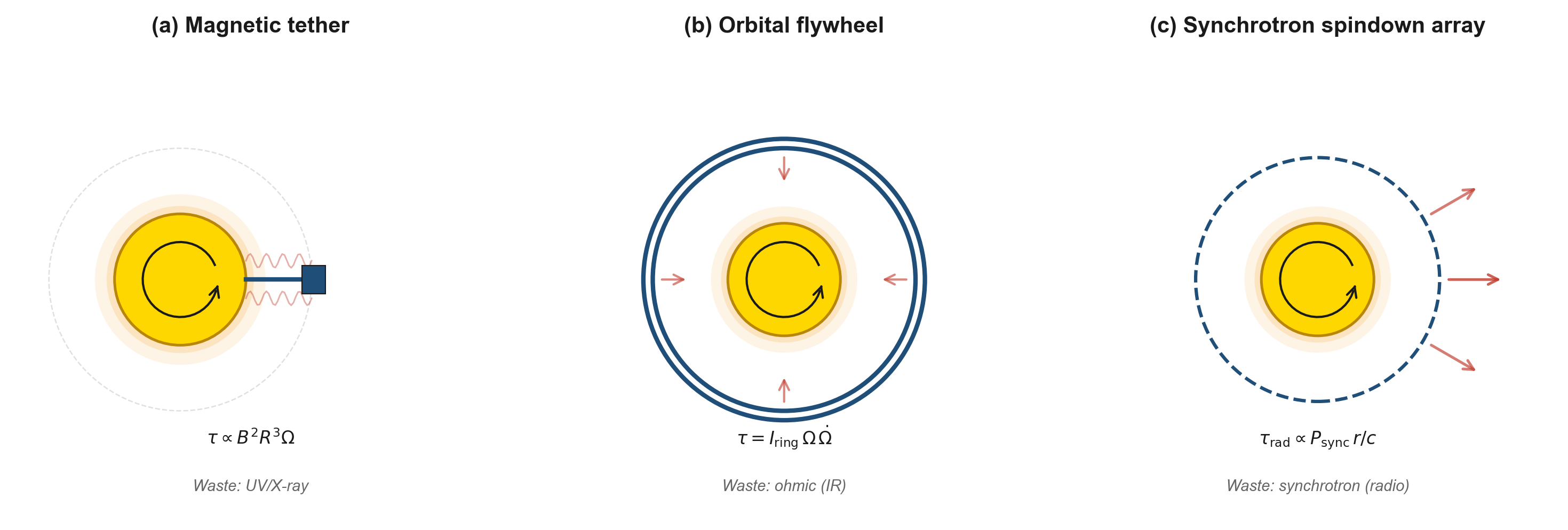}
\caption{Three schematic routes for Stellar J-Harvesting:
(a) magnetic tether, (b) orbital flywheel, and (c) synchrotron spindown
array. These examples illustrate possible coupling channels and associated
waste-heat signatures rather than complete engineering designs.}
\label{fig:mechanisms}
\end{figure*}

\begin{enumerate}
\item {Magnetic tether.} A conducting structure embedded in the stellar
wind could extract angular momentum through Alfv\'en-wave coupling, analogous
to an artificial enhancement of magnetic braking. The torque would scale with
the stellar magnetic field, radius, and rotation rate, schematically
$\tau \propto B_\star^2 R_\star^3 \Omega_\star$. Waste heat would be expected
mainly as ohmic dissipation and possibly chromospheric UV/X-ray emission.

\item {Orbital flywheel.} A massive ring or flywheel could exchange
angular momentum with the stellar magnetosphere through Lorentz-force
coupling. For an illustrative scale, a ring of order
$M_{\rm ring} \sim 10^{22}$\,kg at $r = 1$\,AU
\citep[roughly the habitable-zone distance for FGK hosts;][]{Kasting1993}
could produce percent-level spin-down over Gyr timescales in optimistic
coupling regimes. The dominant waste heat would be ohmic, potentially at
$T \sim 200$--$400$\,K.

\item {Synchrotron spindown array (SSA).} A distributed conducting array
in the stellar wind could accelerate charged particles and emit anisotropic
synchrotron radiation. If the radiation field carries angular momentum away
from the star, the back-reaction torque would oppose stellar rotation. Unlike
the other examples, this route predicts a possible radio excess as a secondary
technosignature, testable with VLA or LOFAR follow-up.
\end{enumerate}
\section{Detection method}
\label{sec:method}

\subsection{A population-relative test that does not need stellar ages}

The traditional approach to identifying anomalous rotators is
gyrochronology: estimate a star's age, predict its expected period, and
flag large deviations. But when the age itself is derived from the
rotation period, the method becomes circular. Rather than untangle this
loop, we sidestep it entirely.

We instead use a local comparison: does a star rotate unusually slowly
relative to stars with similar colour and surface gravity? A strong outlier
in the same $(bp\text{-}rp,\log g)$ cell is then flagged for follow-up
vetting, without first assigning it a gyrochronological age.

The procedure is straightforward:
\begin{enumerate}
\item Bin all stars by $(bp\text{-}rp, \log g)$ in 0.1\,dex cells,
where $\log g$ comes from Berger et al.\ (2020) isochrone fits.
\item In each cell, compute the median $\log_{10} P_{\rm obs}$ and
standard deviation $\sigma_{\log P}$.
\item For each star, compute
$\Delta\log P_{\rm pop} = \log_{10} P_{\rm obs} - \langle\log_{10}
P_{\rm obs}\rangle_{\rm cell}$ and
$z_{\rm pop} = \Delta\log P_{\rm pop} / \sigma_{\log P}$.
\end{enumerate}

\subsection{Eight false-positive filters}

Slow rotation has many natural causes. We therefore apply eight filters
to remove obvious astrophysical false positives:

\begin{enumerate}
\item \emph{FGK only:} $4000 \le T_{\rm eff} \le 7000$\,K. M dwarfs
have different dynamo physics; A/F stars lack deep convection zones.
\item \emph{Main sequence:} $\log g \ge 4.2$ (Berger isochrone). Subgiants
expand and slow down naturally via angular momentum conservation.
\item \emph{Pre-TAMS:} isochrone age $<$ terminal-age main sequence.
Post-MS stars are excluded.
\item \emph{Metallicity:} $[\mathrm{Fe/H}] \ge -0.5$. Metal-poor stars
experience weakened magnetic braking and rotate slowly by nature.
\item \emph{No Gaia companion:} $N_{\rm comp} = 0$. No Gaia-detected
companion within $4''$.
\item \emph{No binary correction:} ${\rm KsCorr} = {\rm NaN}$. Stars
flagged for binary correction in Berger (2020) are excluded.
\item \emph{Astrometric:} ${\rm RUWE} < 1.2$. Strict threshold for
astrometric binaries.
\item \emph{Photometric variability:} $R_{\rm per} < 5000$\,ppm. High
variability may indicate eclipsing, ellipsoidal, or heartbeat binaries
\citep{Jayasinghe2021}.
\end{enumerate}

\subsection{Data}

We draw rotation periods from the catalogue of
34{,}030 Kepler stars \citep{McQuillan2014}, which also provides
$T_{\rm eff}$, $\log g$, $P_{\rm rot}$, and $R_{\rm per}$. Stellar
parameters come from the Berger et al.\ (2020) Gaia--Kepler
catalogue \citep{Berger2020}, which provides isochrone-derived $\log g$,
$T_{\rm eff}$, age, and TAMS for 186{,}301 stars (Table~2), plus RUWE,
$N_{\rm comp}$, ${\rm KsCorr}$, and $[\mathrm{Fe/H}]$ (Table~1).

\begin{figure}
\centering
\includegraphics[width=\columnwidth]{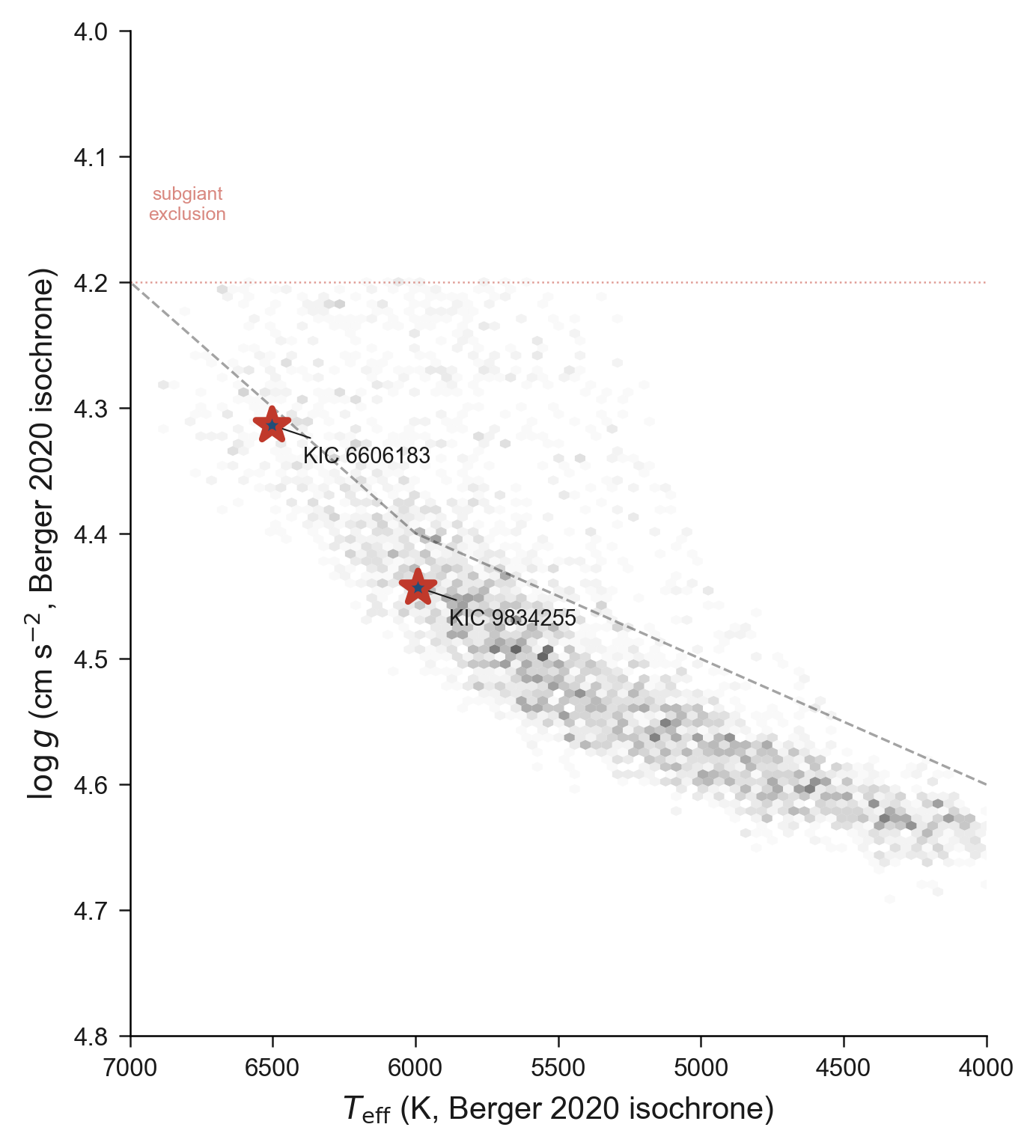}
\caption{The 6{,}725 clean sample stars (grey) in Berger et al.\ (2020)
isochrone $T_{\rm eff}$ versus $\log g$ space. The two $>4\sigma$ SJH
candidates (navy stars) sit firmly on the main sequence, well above the
subgiant exclusion threshold ($\log g = 4.2$, dotted line).}
\label{fig:hr}
\end{figure}

The three-way cross-match yields 32{,}039 stars. After all eight
filters, 6{,}725 clean stars remain (Fig.~\ref{fig:hr}).

\section{Results}
\label{sec:results}

\subsection{Outlier statistics}

Table~\ref{tab:stats} summarizes the slow-rotator tail after all eight
filters. No star exceeds $5\sigma$; two stars remain above $4\sigma$ and
are examined individually below.
\begin{table}
\centering
\caption{SJH outlier statistics after 8 filters
($N_{\rm clean} = 6{,}725$).}
\label{tab:stats}
\begin{tabular}{lrr}
\toprule
Threshold & $N$ & $\Delta\log P_{\rm pop}$ range \\
\midrule
$z > 5\sigma$ & 0 & $\cdots$ \\
$z > 4\sigma$ & 2 & $0.65$ to $1.14$ \\
$z > 3\sigma$ & 6 & $0.55$ to $1.14$ \\
$z > 2\sigma$ & 23 & $0.33$ to $1.14$ \\
\bottomrule
\end{tabular}
\end{table}

\subsection{Two $>4\sigma$ candidates}

Table~\ref{tab:candidates} presents the two highest-significance
candidates. Both are G-type main-sequence stars with isochrone ages of
1.0 and 2.0\,Gyr, well below their TAMS of 4.2 and 8.3\,Gyr. Their
rotation periods of 61 and 65\,d stand in stark contrast to the
5 to 10\,d expected for F/G stars of that age. The implied energy extraction fraction $f_E = 1 - 10^{-2\Delta\log P}$ exceeds 95\%, so an SJH interpretation
would require very strong angular-momentum extraction. We emphasize that
these are candidates requiring spectroscopic and high-resolution imaging
follow-up, not detections.

\begin{table*}
\centering
\caption{Two $>4\sigma$ SJH candidates. Stellar parameters from
Berger et al.\ (2020) isochrone fits; $[M/H]$ and astrometric excess noise
significance from Gaia DR3 \citep{GaiaDR3}. Both candidates have NaN
$[\mathrm{Fe/H}]$ entries in Berger et al.\ (2020) that bypassed our nominal
metallicity filter.}
\label{tab:candidates}
\begin{tabular}{lrrrrrrrrrrr}
\toprule
KIC & $P_{\rm obs}$ & $T_{\rm eff}$ & $\log g$ & Age & TAMS & RUWE & $N_{\rm comp}$ & $[M/H]$ & ${\rm sig}_{\rm EN}$ & $\Delta\log P$ & $z$ \\
 & (d) & (K) & & (Gyr) & (Gyr) & & & & & & \\
\midrule
6606183 & 60.97 & 6504 & 4.31 & 1.04 & 4.17 & 0.99 & 0 & $-0.60$ & 9.4 & 1.14 & 4.12 \\
9834255 & 65.16 & 5992 & 4.44 & 1.97 & 8.27 & 1.09 & 0 & $-0.18$ & 0.0 & 0.65 & 4.12 \\
\bottomrule
\end{tabular}
\end{table*}

\begin{figure}
\centering
\includegraphics[width=\columnwidth]{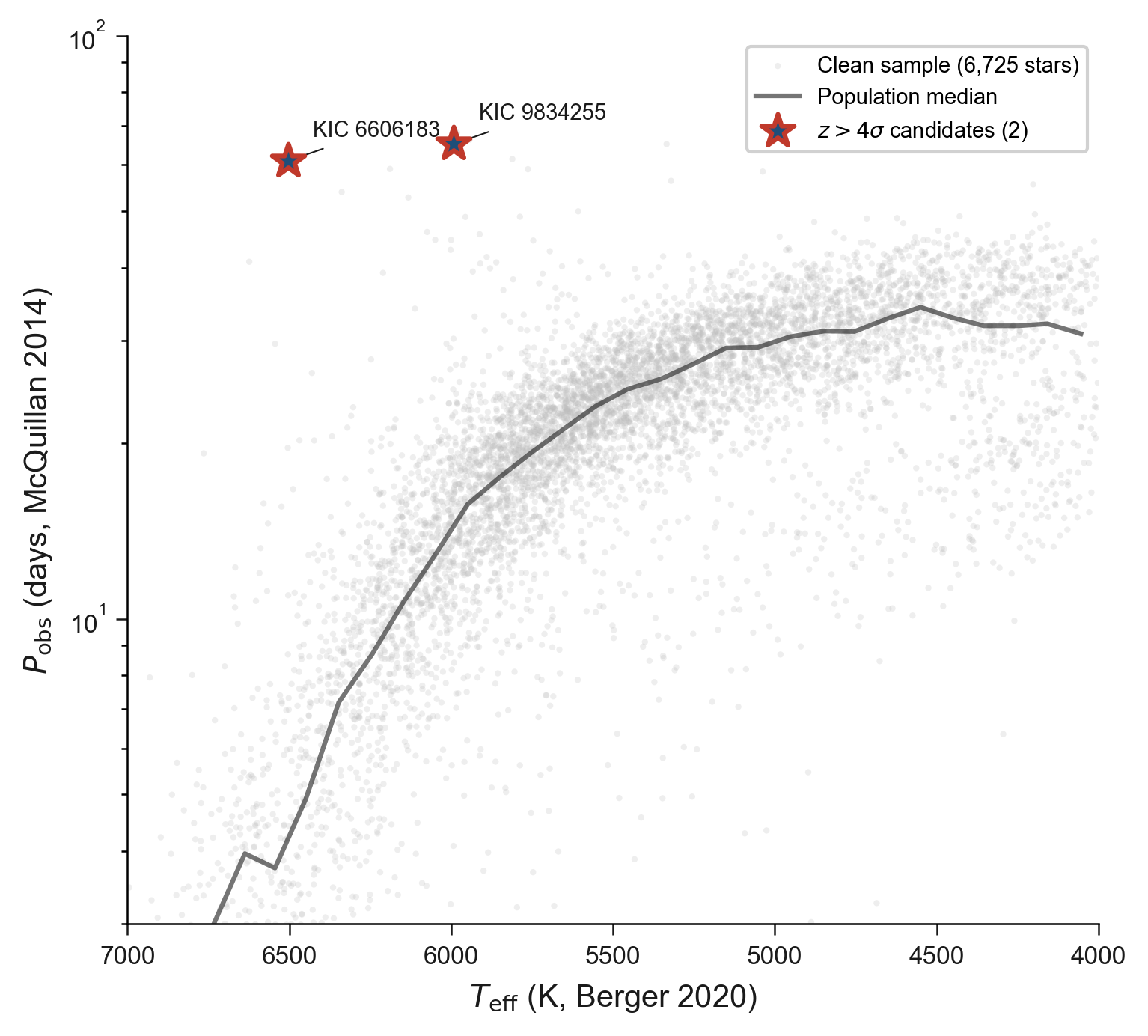}
\caption{Rotation period versus effective temperature for the clean
sample (grey points). The solid line shows the population median in
each $T_{\rm eff}$ bin. The two $>4\sigma$ SJH candidates (navy stars)
sit far above the median, with periods of 61 and 65\,d where 5 to
10\,d would be expected.}
\label{fig:cpd}
\end{figure}

Both candidates formally survive all eight false-positive filters, but
this survival is partly an artifact of missing data: Berger et al.\ (2020)
report NaN $[\mathrm{Fe/H}]$ for both stars, so the metallicity filter does not
actually apply to them. Gaia DR3 \citep{GaiaDR3} fills the gap with
photometric $[M/H] = -0.60$ for KIC\,6606183 and $-0.18$ for
KIC\,9834255. The Gaia astrometric excess noise significance, ${\rm sig}_{\rm EN}$,
also flags KIC\,6606183 with a 9.4$\sigma$ astrometric anomaly that
points to an unseen companion despite its RUWE $= 0.99$. Their
photometric variability ($R_{\rm per} < 800$\,ppm) shows no eclipse-like
signal. Fig.~\ref{fig:cpd} places them in the context of the full
sample.

\subsection{WISE and Aladin images: ambiguous but informative}

\begin{figure*}
\centering
\includegraphics[width=0.9\textwidth]{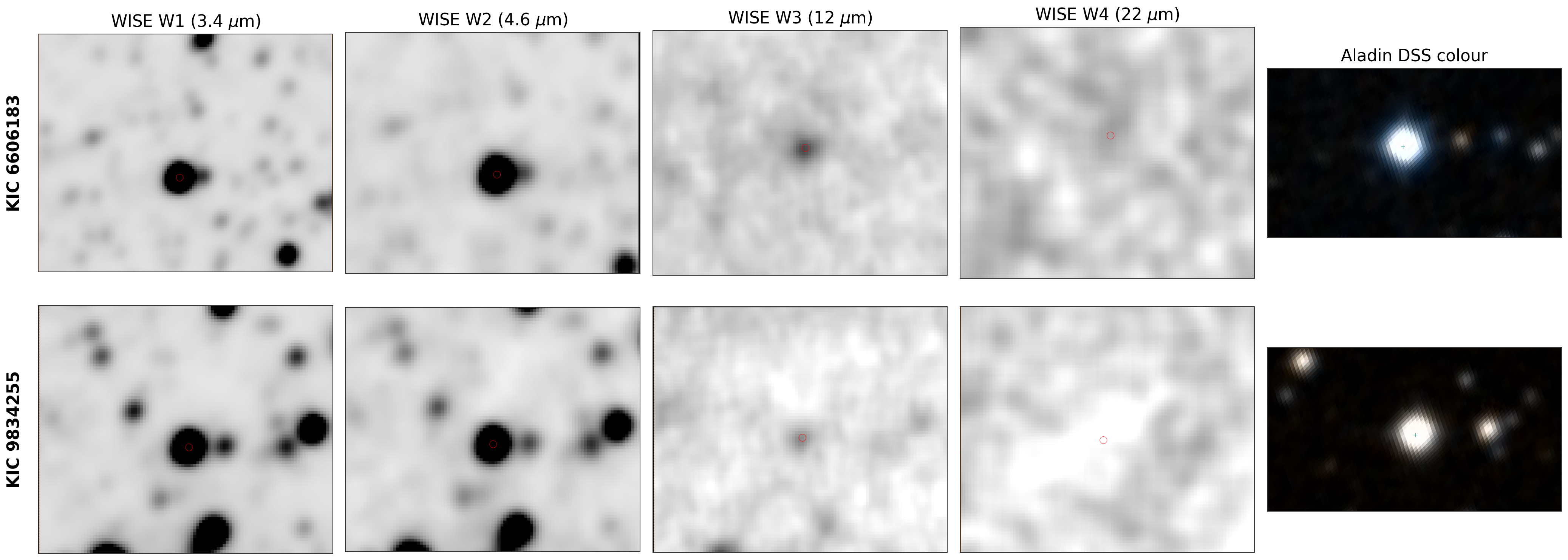}
\caption{Multiwavelength images of the two $>4\sigma$ SJH candidates.
Top: KIC\,6606183. Bottom: KIC\,9834255. Columns: WISE W1--W4 and
Aladin DSS colour composite. Nearby sources are visible but their
physical association requires high-resolution imaging.}
\label{fig:wise}
\end{figure*}

Visual inspection of WISE images (all four bands: W1, W2, W3, W4) and
Aladin optical/infrared overlays reveals infrared sources near both
candidates (Fig.~\ref{fig:wise}). The WISE beam ($\sim$6$''$ at W3,
$\sim$12$''$ at W4) is too coarse to determine whether these are
physical companions or unrelated background objects. Gaia DR3 \citep{GaiaDR3} reports $N_{\rm comp}=0$ for both, but the WISE
beam is too broad to distinguish unrelated background sources from
unresolved companions. High-resolution imaging is required.

No mid-infrared excess attributable to circumstellar dust is detected.
This is consistent with either binary contamination (no SJH signal
expected) or SJH mechanisms whose waste-heat output falls below WISE
sensitivity.

\subsection{Occurrence rate upper limit}

Because both $>4\sigma$ candidates carry compelling binary or
metallicity alternatives (Section~\ref{sec:discussion}), we treat them
as non-detections for the purpose of an occurrence rate upper limit.
With zero candidates at $z > 5\sigma$ in 6{,}725 clean stars, the
Poisson 95\% upper limit \citep[$k=0$;][]{Gehrels1986} is
\begin{equation}
f_{\rm SJH} < \frac{3.0}{6{,}725} \approx 4.5 \times 10^{-4}.
\end{equation}

\section{Discussion}
\label{sec:discussion}

\subsection{Six natural explanations}

\begin{itemize}
\item \emph{Subgiant contamination.} Subgiants rotate slowly because
they expand. An earlier version of this search, using McQuillan
photometric $\log g$, produced 7 candidates at $z > 3\sigma$. Switching
to Berger isochrone $\log g$ immediately identified one as a subgiant
($\log g = 4.03$) and excluded it.

\item \emph{Old-age spin-down.} Old stars rotate slowly; this is the
core of the Skumanich \citep{Skumanich1972} relation. But the Berger
isochrone ages for our two $>4\sigma$ candidates are 1.0 and 2.0\,Gyr,
well below their TAMS. A 1\,Gyr F/G star should rotate at 5 to 10\,d,
not 61 to 65\,d. The old-age explanation does not survive the age
data.

\item \emph{Binary tidal braking.} This is the most stubborn
false-positive channel and the most likely explanation for both
surviving candidates. Of the original 7 candidates at $z > 3\sigma$
(before binary filters), 3 were excluded: two by $N_{\rm comp} > 0$ and
KsCorr binary flags, one by ${\rm RUWE} = 1.55$. The survivors have
${\rm RUWE} < 1.1$ and $N_{\rm comp} = 0$, but close spectroscopic
binaries (SB1) or sub-arcsecond binaries can escape both Gaia
astrometric detection and the KsCorr flag. Stellar multiplicity is
common among solar-type stars \citep{Kraus2007}. The Gaia DR3
\citep{GaiaDR3} astrometric excess noise significance for KIC\,6606183
is 9.4$\sigma$, a strong astrometric binary indicator that the RUWE
threshold misses. KIC\,9834255 has 6 blended BP transits in Gaia DR3,
suggesting a nearby source within the BP aperture. Both stars therefore
carry independent binary indicators that survived our filters, and we
cannot exclude this channel without radial-velocity monitoring and
high-resolution imaging.

\item \emph{Weakened magnetic braking.} The classical rotation--activity
relation \citep{Noyes1984} underpins the Rossby-number framework, and
\citet{vanSaders2016} showed that braking weakens at $R_o \gtrsim 2$.
Both candidates have $R_o < 1.1$ (computed via \citealt{CranmerSaar2011}
convective turnover times), ruling out this channel.

\item \emph{Metallicity.} Low metallicity weakens magnetic braking and
produces naturally slow rotation. Both candidates have NaN $[\mathrm{Fe/H}]$
entries in Berger et al.\ (2020), so our nominal $[\mathrm{Fe/H}] \ge -0.5$
filter did not apply to them. Gaia DR3 \citep{GaiaDR3} photometric
metallicities fill this gap: $[M/H] = -0.60$ for KIC\,6606183 and
$[M/H] = -0.18$ for KIC\,9834255. The first value sits below our nominal
$-0.5$ cutoff, meaning KIC\,6606183 would have been excluded had the
metallicity been available, and its slow rotation is at least partially
explained by weakened magnetic braking at low $[M/H]$ without invoking
SJH. KIC\,9834255 is near solar, so the metallicity channel is less
convincing for it, but binary contamination remains. Spectroscopic
$[\mathrm{Fe/H}]$ from LAMOST or APOGEE is essential for both candidates.

\item \emph{Kraft break.} KIC\,6606183 ($T_{\rm eff} = 6504$\,K) sits
near the Kraft break ($\sim$6250\,K) where the convective envelope
thins and magnetic braking becomes less efficient; we flag this as a
systematic that could inflate its $z$-score. KIC\,9834255
($T_{\rm eff} = 5992$\,K) lies well below the break with a deep
convective envelope, so this concern does not apply to it.
\end{itemize}
\subsection{Where does the waste heat go?}

Thermodynamics demands that SJH produce waste heat. But the total power
($\dot{E} \sim 10^{18}$\,W for 50\% extraction over 1\,Gyr) is
$\sim$$10^8$ times fainter than a Dyson sphere intercepting the full
stellar luminosity. At typical Kepler distances ($\sim$1\,kpc), the
resulting flux falls below the WISE detection threshold by several
orders of magnitude. SJH is, by design, a stealth technosignature:
invisible to infrared surveys, detectable only through its rotational
signature.

The SSA method (Section~\ref{sec:concept}) offers a unique exception. Its synchrotron
waste heat is emitted in the radio band, making it detectable with
VLA or LOFAR observations. This gives us an independent search channel
that no previous technosignature survey has explored.

\subsection{Limitations}
\begin{enumerate}
\item \emph{Coverage:} 6{,}725 stars is a very small fraction of the Gaia DR3 \citep{GaiaDR3}
FGK sample. Gaia DR4 will provide an extensive data set that will improve data coverage.
\item \emph{Single catalogue:} All periods come from McQuillan (2014),
and the sample inherits Kepler's geometric selection effects
\citep{Stevens2013}. Independent verification from TESS
\citep{Reinhold2020} or ZTF is needed.
\item \emph{No spectroscopic follow-up:} $v \sin i$, radial velocity,
and chromospheric activity ($\log R'_{\rm HK}$) are required to
validate candidates and test the rotation--activity decoupling predicted
for SJH-braked stars.
\item \emph{WISE resolution:} The $6''$ beam cannot resolve
sub-arcsecond companions. High-resolution imaging is essential.
\item \emph{Age scatter:} The population-relative test does not fully
control for age variation within a $(bp\text{-}rp, \log g)$ cell.
Berger isochrone ages help (candidates are young), but residual scatter
remains.
\end{enumerate}
Beyond the present Kepler pilot sample, future SJH searches could combine
slow-rotator outlier ranking with habitability-driven SETI target-selection
filters, such as the rule-based Gaia DR3 avoidance framework of
\citet{Torlakcik2026}. This would separate two questions: whether a
star is rotationally anomalous, and whether it is a high-priority SETI
target for follow-up.

\section{Conclusions}
\label{sec:conclusions}

We have introduced Stellar J-Harvesting as a new
angular-momentum-based technosignature, complementary to the canonical
Dyson search. The energy--period relation gives a direct link between
angular-momentum extraction and anomalously slow stellar rotation. The
engineering mechanisms discussed here are illustrative rather than complete
designs, but they show plausible coupling channels. The main challenge is
not defining the observable, but separating rare artificial spin-down from
the natural slow-rotation channels already present in stellar catalogues.

Using 6{,}725 clean FGK main-sequence stars from McQuillan et al.\
(2014) and Berger et al.\ (2020), filtered through eight
false-positive channels, we identify 2 candidates at $z > 4\sigma$:
KIC\,6606183 ($P = 61$\,d, age $= 1.0$\,Gyr) and KIC\,9834255
($P = 65$\,d, age $= 2.0$\,Gyr). If interpreted as SJH signals, their periods would imply $>95\%$ energy extraction. WISE imaging
reveals nearby sources whose nature remains ambiguous.

We do not claim a detection. We report these as candidates requiring
spectroscopic and high-resolution imaging follow-up, and we place a
conservative upper limit $f_{\rm SJH} < 4.5 \times 10^{-4}$. Both
candidates have compelling astrophysical alternatives: KIC\,6606183 has
Gaia DR3 $[M/H] = -0.60$ (below our nominal metallicity cutoff) and a
9.4$\sigma$ astrometric excess noise significance, while KIC\,9834255
has 6 blended BP transits indicative of a nearby source. Binary
contamination is the dominant false-positive channel, and future
searches must incorporate radial-velocity monitoring and
high-resolution imaging to progress beyond this preliminary
characterization.

The broader implication is that technosignature space is not limited to
infrared waste heat. A civilization harvesting angular momentum could leave
its clearest trace not in luminosity, but in stellar rotation.

\section*{Acknowledgements}
This work was carried out independently and received no external funding. This work has made use of Kepler, McQuillan et al.\ (2014)
rotation period data, and Berger et al.\ (2020) Gaia--Kepler stellar
properties catalogue. Software: \texttt{numpy}, \texttt{pandas},
\texttt{astropy}, \texttt{scipy}. The author used Gemini for language editing and readability improvements. The scientific analysis, interpretation, and conclusions are entirely the author's own.

\section*{Data Availability}

The McQuillan et al.\ (2014) rotation period catalogue and the
Berger et al.\ (2020) Gaia--Kepler stellar properties catalogue are
publicly available via VizieR. Gaia DR3 data are available through the
Gaia Archive. The candidate list and pipeline underlying
this article are available from the author upon reasonable request.

\bibliographystyle{mnras}
\bibliography{bibliography}

\label{lastpage}
\end{document}